\documentclass[usenatbib]{mn2e}
\usepackage{epsfig}
\begin{document}
 
 \title[Cosmological parameters from the SFI++ survey]{Peculiar velocities into the next generation: cosmological parameters from the SFI++ survey}

\author[Abate and Erdo{\u g}du]{Alexandra Abate$^1$ and Pirin Erdo{\u g}du$^{2,3}$\\
$^1$Laboratoire de l'Acc\'{e}l\'{e}rateur Lin\'{e}aire, IN2P3-CNRS, Universit\'{e} de Paris-Sud, BP. 34, 91898 Orsay Cedex, France\\
$^2$Department of Physics and Astronomy, University College London, London, WC1E 6BT\\
$^3$Department of Science and Engineering, American University of Kuwait, P.O. Box 3323, Safat 13034, Kuwait}

\maketitle

\begin{abstract}
We present cosmological parameter constraints from the SFI++ galaxy peculiar velocity survey, the largest galaxy peculiar velocity sample to date.  The analysis is performed by using the gridding method developed in \cite{abate6df}.  We concentrate on constraining parameters which are affected by the clustering of matter: $\sigma_8$ and the growth index $\gamma$.  Assuming a concordance $\Lambda$CDM model we find $\sigma_8=0.91^{+0.22}_{-0.18}$ and $\gamma=0.55^{+0.13}_{-0.14}$ after marginalising over $\Omega_m$.  These constraints are consistent with, and have similar constraining power to, the same constraints from other current data sets which use different methods.   Recently there have been several claims that the peculiar velocity measurements do not agree with $\Lambda$CDM.  We find instead although a higher value of $\sigma_8$ and a lower value of $\Omega_m$ are preferred, the values are still consistent when compared with WMAP5.  We note that although our analysis probes a variety of scales, the constraints will be dominated by the smaller scales, which have the smallest uncertainties. These results show that peculiar velocity analysis is a vital probe of cosmology, providing competitive constraints on parameters such as $\sigma_8$.  Its sensitivity to the derivative of growth function, particularly down to redshift zero, means it can provide a vital low redshift anchor on the evolution of structure formation. The importance of utilising different probes with varying systematics is also an essential requirement for providing a consistency check on the best-fitting cosmological model.

\end{abstract}

\section{Introduction}
\renewcommand{\thefootnote}{\fnsymbol{footnote}}
\setcounter{footnote}{1}
\footnotetext{E-mail: abate@lal.in2p3.fr}

The cosmological velocity field is induced by the gravitational affect of the inhomogeneities in the matter distribution in the universe.  Therefore the recessional velocities of galaxies experience deviations from pure Hubble flow, the result of the expansion of space, due to the distribution of matter both luminous and dark.  The amplitude and distribution of these deviations or \textit{peculiar velocities} allows measurement of the power spectrum of the fluctuations in the matter distribution.  This is an important quantity to constrain because it tells us how structure has grown on different scales in the universe, and this in turn depends on the amount of and nature of the components in the universe, i.e. the baryonic matter, dark matter, dark energy.

The distribution of galaxies in the universe is not likely to be the same as the distribution of matter, since most of the mass is in the form of indirectly detectable dark matter.  It is known that galaxies of different types cluster differently \citep{dresslerbias,bias}, so they are clearly not completely unbiased tracers of the underlying mass, and this issue is referred to generically as galaxy biasing.  Comparatively therefore, the distribution of galaxy peculiar velocities has an advantage over the distribution of galaxies in probing the matter distribution because they are likely to be unbiased traces of the matter velocity field. This is then simply related to the density field in linear theory.  Since peculiar velocities are a non-local function of the dark matter distribution then analysing the peculiar velocity field provides information on scales larger than the sampled region \citep{hoff} as the velocity at a point is determined by the integral over the matter distribution in a large volume.

Peculiar velocities are interesting for another reason as they provide the only way to measure clustering of objects at practically redshift zero,  
noting that the clustering measured by the density field is complicated by redshift space distortions and galaxy bias.
Weak lensing, Lyman-$\alpha$ forest (Ly-$\alpha$) and cluster count measurements are obtained only at higher redshifts, often where dark energy was just starting to dominate. Understanding the strength and amount of clustering in the local universe could help discriminate between different models that affect the growth of large scale structure such as dark energy models and modified gravity.

In practice peculiar velocities are complicated by several factors. The major one is that on small scales the density field is highly nonlinear; these effects leak into the velocity field and cannot be described analytically.  Results from previous surveys which apply likelihood analysis \cite[see][for Mark III, ENEAR and SFI respectively]{zab97,zab,freud} seemed to overestimate the combination $\sigma_8\Omega_m^{0.6}$ significantly compared to other probes at the time and to the current concordance cosmology. Studies by \cite{hoffzar} and \cite{silb} show that this over-estimation may be due to inaccurate modelling of the nonlinear part of the power spectrum, the small scales. The method employed in this paper overcomes this issue by smoothing the small-scale velocity field by averaging together peculiar velocity measurements regularly in a small volume of space, see Section \ref{section:grid} for details.  

Another disadvantage of peculiar velocity analysis is the accuracy of the peculiar velocity measurements. Because a redshift-independent estimate of the distance to the galaxy is required, and these estimates are usually extremely noisy.  Most previous peculiar velocity surveys have used distance indicators such as the Fundamental Plane/Dn-$\sigma$ \citep{fp,dnsig} and the Tully-Fisher relation \citep{tf}. These are galaxy scaling relations and it is because of their large scatter that the distance estimates are so noisy.  The calculated distance is a relative distance measure which is strongly subject to a number of biases and also has very large uncertainties of around 20 percent, all of which translates to the peculiar velocity \cite[see][for a review]{S&W}.  Although type 1a supernovae (SN1a) are more accurate as distance indicators (they have smaller distance errors, 5\% to 10\%) the current samples are very small (around 150), whereas galaxy peculiar velocity samples are always in excess of 1000 objects.  The SFI++ \citep{sfi++1,sfi++} is the largest galaxy sample to date so provides an excellent opportunity to improve the statistics of peculiar velocity surveys.

There has recently been a resurgence of interest in peculiar velocity analysis, after it fell out of favour in the late 1990s due to the apparent overestimation of $\sigma_8\Omega_m^{0.6}$ described above.  The renewed interest can be roughly catagorised into the following areas: bulk flow measurements; as a ``useful" systematic to SN1a luminosity distance measurements; the kinematic Sunyaev-Zeldo'vich (kSZ) effect.  Bulk flow measurements have been recently made by \cite{danish07}, \cite{GLS08}, \cite{fw08} and \cite{wfh09}.  Measurements of the large scale velocity field, or bulk flow, has a number of interesting applications. If a difference exists between measurements of our dipole motion made using local peculiar velocities, and the dipole motion inferred from the cosmic microwave background (CMB) then this may be an indication of a departure from the standard WMAP5 cosmological model \citep[as published in][]{wmap5}.  Otherwise, because the bulk flow is a sensitive probe of the matter density fluctuations on very large scales, it can probe the amplitude and shape of the matter power spectrum.  


Two recent studies of peculiar velocity surveys indicated a large bulk flow. \cite{kash} found a strong bulk flow on scales out to 300 $h^{-1} {\rm Mpc}$ using the kSZ. \cite{wfh09} reported a large-scale bulk flow beyond 50 $h^{-1} {\rm Mpc}$ from a comprehensive compilation of peculiar velocity data. Both these papers claimed that the bulk flow at these depths, determined by their studies, were difficult to explain within the framework of standard $\Lambda$CDM model of cosmology. 

It has been recently found that the peculiar velocity field does in fact affect measurements of the expansion of the universe from SN1a.  Low redshift (0.02$<z<$0.1) SN1a are required both as a well-observed calibrating sample and as a lever arm for the SN1a Hubble diagram \cite[see][for more details]{NSF,abatesn1a}. Although we are more interested in the expansion history of the universe at redshifts $>0.2$, these low redshift ''calibrating" SN1a have sufficiently low redshifts that the local velocity field systematically correlates their luminosity distance measurements.  This effect must either be corrected for \citep[e.g.][]{NHC07} or modelled \citep[e.g.][]{Hui06,Cooray06,GLS07} to retrieve an unbiased measurement of the dark energy equation of state $w$.

The paper is organised as follows, in Section \ref{section:data} we describe the SFI++ data set, and in Section \ref{section:meth} we describe our method of analysing this data. Our results are presented in Section \ref{section:res} and discussion and conclusions of these results are in sections \ref{section:disc} and \ref{section:conc}.

\section{SFI++ Peculiar Velocity Data}
\label{section:data}
The SFI++ sample \citep{sfi++1} consists of around 5000 spiral galaxies that have measurements which are suitable for the application of the I-band Tully-Fisher (TF) relation. This sample builds on the older SCI and SFI samples, but includes significant amounts of new data, as well as improved methods for parameter determination.  The rotation widths for the SFI++ sample come from the 21-cm line global profile widths (HI, about 60\%) and optical rotation curves (H$\alpha$, about 40\%).  The catalogue presented in \cite{sfi++} represents a homogeneously derived set of distances and peculiar velocities that are corrected for both the homogeneous and inhomogeneous Malmquist biases using the 2MASS redshift survey \citep{2masshuc} for the line-of-sight density field.

The data set we use comes from \cite{sfi++err} after some corrections were made to the original tables published in \cite{sfi++}.  The morphological-type correction for the peculiar velocity calculation was not properly applied in the original tables and this problem has now been rectified.  The data set contains 4053 galaxies which have a median redshift of 0.019\footnote{The data set can be downloaded from: http://www.iop.org/EJ/article/0067-0049/182/1/474/}.  The magnitude of the error on the inferred distance is approximately 22\% of the distance estimate. The sample covers most of the sky above the Galactic plane, with some deficiency of galaxies in the declination range of $\delta= [-17.5^{\circ},-2.5^{\circ}]$. 

\section{Methodology}
\label{section:meth}
The analysis described in this Section follows \cite{abate6df}, for full details please refer to that paper. 
We use linear theory to predict the velocity correlation function and use a multivariate Gaussian to calculate the likelihood.  Section \ref{section:grid} describes how we overcome two potential problems: biases from nonlinear growth of structure and the large number of velocities in the survey.  
See also the work by \cite{jaffe},  
who perform a similar cosmological likelihood analysis using a Friends-of-Friends technique.

\subsection{Likelihood analysis using the VCF}
\label{subsection:vcf}
To estimate cosmological parameters we compute the radial peculiar velocity correlation function (hereafter VCF) from linear theory for each set of parameters. We define the peculiar velocity with observational error as $\textbf{v}_i \cdot \hat{\textbf{r}}_i\equiv v_i =s_i+\epsilon_i$, therefore the observed VCF is defined by
\begin{eqnarray}
\label{eq:rij}
R_{ij}& =&\left< v_i v_j \right>=\left< s_i s_j \right>+\left<\epsilon_i \epsilon_j\right> \\
                           &=&\xi_{ij}+\epsilon_i^2\delta_{ij}
\end{eqnarray}where the average is over realisations of the universe. In practice we average over a large enough volume of space to assume ergodicity. The first term is the signal VCF and the second term is the contribution from the errors in the velocity measurements. Because the errors are assumed to be uncorrelated so they only affect the diagonal terms in $R_{ij}$.  In linear theory, the signal part $\xi_{ij}$ can be split up into perpendicular and parallel components \citep{gorski,gjo}, which are scalar functions of r=$|\textbf{r}|$ 
\begin{equation}
\label{cf}
\xi_{ij}=\cos \theta_i\cos \theta_j\Psi_{||}(r)+\sin \theta_i\sin \theta_j\Psi_{\perp}(r)
\end{equation} where the angles are defined by $\cos\theta_{X}=\hat{\textbf{r}}_{X} \cdot \hat{\textbf{r}}$ and the diagonal elements $\xi_{ii}$ are given by Eq.~\ref{eq:diag} below. The $\Psi_{||}(r)$ and $\Psi_{\perp}(r)$ represent the velocity correlations along the line-of-sight and perpendicular to the line-of-sight respectively.
They are calculated from the matter power spectrum and assuming all galaxies are approximately at redshift zero
\begin{equation}
\label{eq:psi}
\Psi_{||,\perp}(r)=\frac{H_0^2 f^2(\Omega_m)}{2 \pi^2} \int P(k) B_{||,\perp}(kr) dk
\end{equation}
where $B_{\perp}=j_0^{'}(x)/x$ and $B_{||}=j_0^{''}(x)$ and $ j_0^{'}, j_0^{''}$ are the first and second derivative of the zeroth order spherical Bessel functions respectively, $H_0$ is the Hubble constant and $\Omega_m$ is the density of matter in the universe normalised by the critical density, $f(\Omega_m)$ is  the derivative of the growth function.  The auto correlation is given by
\begin{equation}
\label{eq:diag}
\xi_{ii}=\frac{1}{3}\frac{H_0^2 f^2(\Omega_m)}{2 \pi^2} \int P(k)dk.
\end{equation}
The dependence on the cosmological parameters of interest enters the above equations in the following ways:
$\sigma_8$ through the normalisation of the power spectrum $P(k)$; $\Omega_m$ through f$(\Omega_m)\cong\Omega_m^{\gamma}$ \cite[where $\gamma\simeq0.55$,][]{wangstein} and through its effect on the shape of the matter power spectrum; $\gamma$ through $f(\Omega_m)$ as stated above. To compute Eq.~\ref{eq:psi} and Eq.~\ref{eq:diag} the power spectrum $P(k)$ is generated using CAMB \citep{camb}. 

The above equation for $f$ assumes that the galaxies are at low redshift, therefore equations \ref{eq:psi} and \ref{eq:diag} are only valid for low redshifts. The full equation contains the growth rate at the redshift of each galaxy but because of the redshift range of SFI++ it is unnecessary to do the full computation here.  Peculiar velocity surveys using distance indicator relations (as SFI++ does) are unlikely to have a significant amount of data beyond a redshift of 0.05. The growth rate increases by less than 1 per cent between redshift zero and redshift 0.05 for a flat $\Lambda$CDM model with $\Omega_m=0.3$, so this is a good approximation for this paper. We also use the approximation that the Hubble expansion is constant, described simply by a constant expansion rate for all galaxies in the survey. 

To calculate the covariance matrix $\bf{R}$ (Eq.~\ref{eq:rij} in matrix notation) it is necessary to calculate the equations \ref{eq:rij} to \ref{eq:diag} for a given set of cosmological parameters $\bf\Theta$. Assuming that the peculiar velocities and the observational errors are Gaussian random fields the likelihood function for the parameter set $\bf\Theta$ can be written as
\begin{equation}
\label{eq:like}
\textit{L}(\bf\Theta)=\frac{1}{\sqrt{(2\pi)^N |\bf{R(\bf\Theta)}|}} \exp \left( -\frac{1}{2} \sum_{i, j}^{N} v_i\,\, ({\bf R}^{-1}(\bf\Theta))_{ij}\,\, v_j\right).
\end{equation} where $\bf v_i$ are the observed radial velocities.  

The likelihood analysis outlined above uses the individual galaxy peculiar velocities, $v_i$, as the data.  However determining cosmological parameters in this way does not take account of the nonlinear part of the peculiar velocity signal because, as stated above, we make our prediction for the VCF based only on linear theory.  The density field becomes nonlinear only on small scales, above a wave number ($k$) of about 0.2$h$ Mpc$^{-1}$.  The next section describes how we alter the above analysis to take account of this nonlinear signal.

\subsection{Gridding Method}
\label{section:grid}
This method is a way of averaging together the peculiar velocities of spatially close galaxies by laying a grid across the survey.  Averaging over a number of galaxies allows the linear signal to dominate.  This averaging however will necessarily smooth the velocity field which is equivalent to damping the small scale contributions. This reduces the observed correlations because we average away some of the signal.  If the data is averaged on a grid and inserted directly into the equations in Section~\ref{subsection:vcf} without accounting for the averaging in the correlation function then the cosmological parameters will be biased. 
Therefore we present below how to implement this type of binning on the velocity field, and detail a practical approach to take account of the binning accurately in the VCF. 
We do not follow the approach seen in previous work which includes an effective noise term due to any remaining non-linear signal \cite[e.g.][]{jaffe,zab}.  \cite{zab}  found including such a noise term has only marginal effects on the results, and we have shown in \cite{abate6df} that the grid averaging and the window function given by Eq.~\ref{eq:window} removed any nonlinear bias on $\sigma_8$ to well within the statistical precision.
The technique we implement here is designed to be simple and fast.  For a more complete description of this method, including its performance after rigorous testing with simulations, see \cite{abate6df}.

The method is implemented as follows. We lay down a grid across the survey and average together all the peculiar velocities within each grid cell so that
\begin{eqnarray}
\label{eq:cell}
v_m^{\prime} &=&\left<v_i \right>_{i\in m} \\
\epsilon^{\prime}_{m} &=& \frac{\left< \epsilon_i \right>_{i \in m}}{\sqrt{n_{m}}}
\label{eq:errorcell}
\end{eqnarray}
where $v_m^{\prime}$ is the radial peculiar velocity of the cell and $\epsilon_{m}^{\prime}$ is the error on the velocity of the cell; the angle brackets denote an average over all galaxies $i$ within the cell $m$.  
Note that $\epsilon_i$ is the contribution to the correlation function from the random velocity errors of each galaxy and therefore the remaining contribution $\epsilon^{\prime}_m$ to the binned correlation function is reduced by the square root of the number of galaxies.

This type of binning is then taken account of by multiplying the power spectrum in Eq.~\ref{eq:psi} and~\ref{eq:diag} with a window function corresponding to the size and shape of the grid cell, which has the following form in Fourier space  
\begin{equation}
\label{eq:window}
W(k)=\left< \frac{8}{L^3}\frac{\sin \left(k_x \frac{L}{2}\right)}{k_x}\frac{\sin \left(k_y \frac{L}{2}\right)}{k_y}\frac{\sin \left(k_z \frac{L}{2}\right)}{k_z}
\right>_{{\bf k} \in k}
\end{equation}
where $L$ is the length of a side of a cell in the grid, and the angle brackets denote an average over Fourier space directions.  This means equations that Eq.~\ref{eq:psi} and~\ref{eq:diag} become
\begin{eqnarray}
\label{eqn:sismooth}
\Psi^{\prime}_{||,\perp}(r)&=&\frac{H_0^2 f^2(\Omega_m)}{2 \pi^2} \int W^2(k)P(k) B_{||,\perp}(kr) dk \\
\label{eqn:sidiagsmooth}
\xi^{\prime}_{mm}&=&\frac{1}{3}\frac{H_0^2 f^2(\Omega_m)}{2 \pi^2} \int W^2(k)P(k)dk
\,
\end{eqnarray} where we use the position of the cell centre to calculate all the required distances. The corresponding VCF and likelihood are then formed using the smoothed quantities.

The method for accounting for the velocities described above assumes the data inside each grid cell is a continuous field, whereas it is in fact discrete values at the locations of the galaxies; we shall refer to this as the \textit{sampling effect}.  Galaxies trace discrete points of the peculiar velocity field, but if enough discrete points are averaged over then they will closely approximate averaging over a continuous distribution. However, some grid cells will not contain enough galaxies to provide a reasonable measure of the average of the velocity field within that cell, perhaps due to masked out areas in the survey or poor sampling in some areas. For example there is a deficiency of galaxies in the declination range of $\delta= [-17.5^{\circ},-2.5^{\circ}]$. 
The sampling effect is not simply shot noise.  The under-sampling of the velocity of a cell by a finite number of galaxies causes a spurious boost in the variance
 and if this effect is unaccounted for it could bias the cosmological parameters, e.g. the effect on $\sigma_8$ would be to bias it to too high a value.

We found in \cite{abate6df} if we reduce the size of the modification to the diagonal elements of the VCF, $\xi_{ii}$, according to the number of galaxies in each cell we could remove any bias caused by the sampling effect.  The following relation for this reduction was used:
\begin{equation}
\label{eq:corr}
\xi_{mm}^{{\rm corr}}=\xi_{mm}^{\prime}+\frac{\left(\xi_{mm}-\xi_{mm}^{\prime}\right)}{n_{m}}
\end{equation}
where $\xi_{mm}^{\rm corr}$ is the corrected value used to calculate the diagonal elements of the VCF; $\xi_{mm}^{\prime}$ is calculated from Eq.~\ref{eqn:sidiagsmooth}; $\xi_{mm}$ is calculated from  Eq.~\ref{eq:diag}. This correction uses the correct value for the diagonal elements of the correlation function in the limit that there is just one galaxy in the cell and also in the limit that there are infinite galaxies in the cell, i.e. a continuous field; for intermediate numbers of galaxies the value of diagonal elements of the VCF should be an improvement on using $\xi_{ii}^{\prime}$.  

\section{Results}
\label{section:res}

\begin{figure}
\center
\epsfig{file=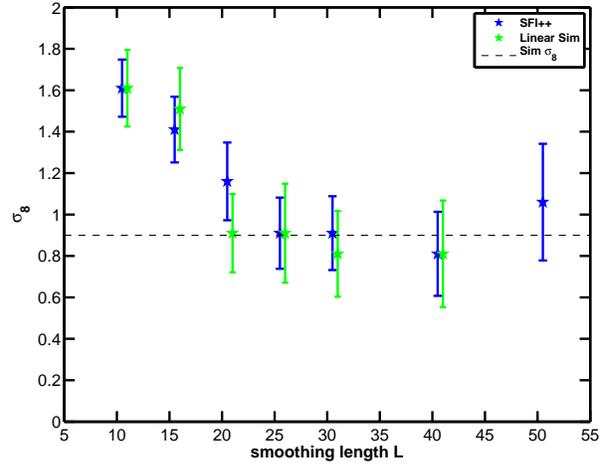,width=8cm,angle=0}
\caption{Best-fit $\sigma_8$ for various smoothing length $L$. The bold (blue) points are the results from the SFI++ data, the faint (green) points are the results from the linear simulation.  The dashed line is the fiducial $\sigma_8$ for the linear simulation. \label{varygrid}}
\end{figure}

We assume the following cosmological parameters throughout: $h=0.7$, $n_s=1$, $\Omega_b=0.05$ and $\Omega_k=0$.

First we need to determine the optimum smoothing length $L$ so as not to introduce bias in the cosmological parameters from nonlinear signal or from the smoothing technique itself.  To do this we find the best-fit $\sigma_8$ for various smoothing length $L$ for both the SFI++ data, and also a linear simulation of around 1500 galaxies.  For this we have additionally fixed $\Omega_m=0.3$.  The results of this are shown in Figure \ref{varygrid}.  The bold (blue) points are the results from the SFI++ data, the faint (green) points are the results from the linear simulation. 

We simulate galaxy velocities for the linear simulation with the correlation function ${\bf \xi}$ in Eq.~\ref{cf}. The number of galaxies in SFI++ is large enough that computation of the covariance matrix and eigenvalues was prohibitively slow so we simulate just 1500 galaxies.  We use real positions of galaxies in SFI++ data set but simulate 50 realisations of peculiar velocities, and the resulting constraint on $\sigma_8$ is the average over these 50 realisations.  The fiducial $\sigma_8$ for the linear simulation was $\sigma_8=0.9$ and is shown on Figure \ref{varygrid} by the horizontal dashed line. 

The large bias at $L<20$Mpc/$h$ is mainly an artifact of the gridding method, and not due in any major way to nonlinear bias or other systematic.  This is confirmed because the linear simulation results follow closely the results of the SFI++ data, and recover the fiducial $\sigma_8$ after $L=20$Mpc/$h$.  Therefore we find that $L=25$Mpc/$h$ is the best choice of smoothing length: large enough to avoid any bias but small enough so as not to redundantly increase the statistical error on the cosmological parameters.  This value is close to the value of the optimum smoothing length found in \cite{abate6df} of $L=20$Mpc/$h$.
To check we are not affected by velocity bias, we repeated the linear simulation, with the same number of objects but with uniform distribution instead of a distribution which matched SFI++.  We did not find any significant difference in the results.

\begin{figure}
\center
\epsfig{file=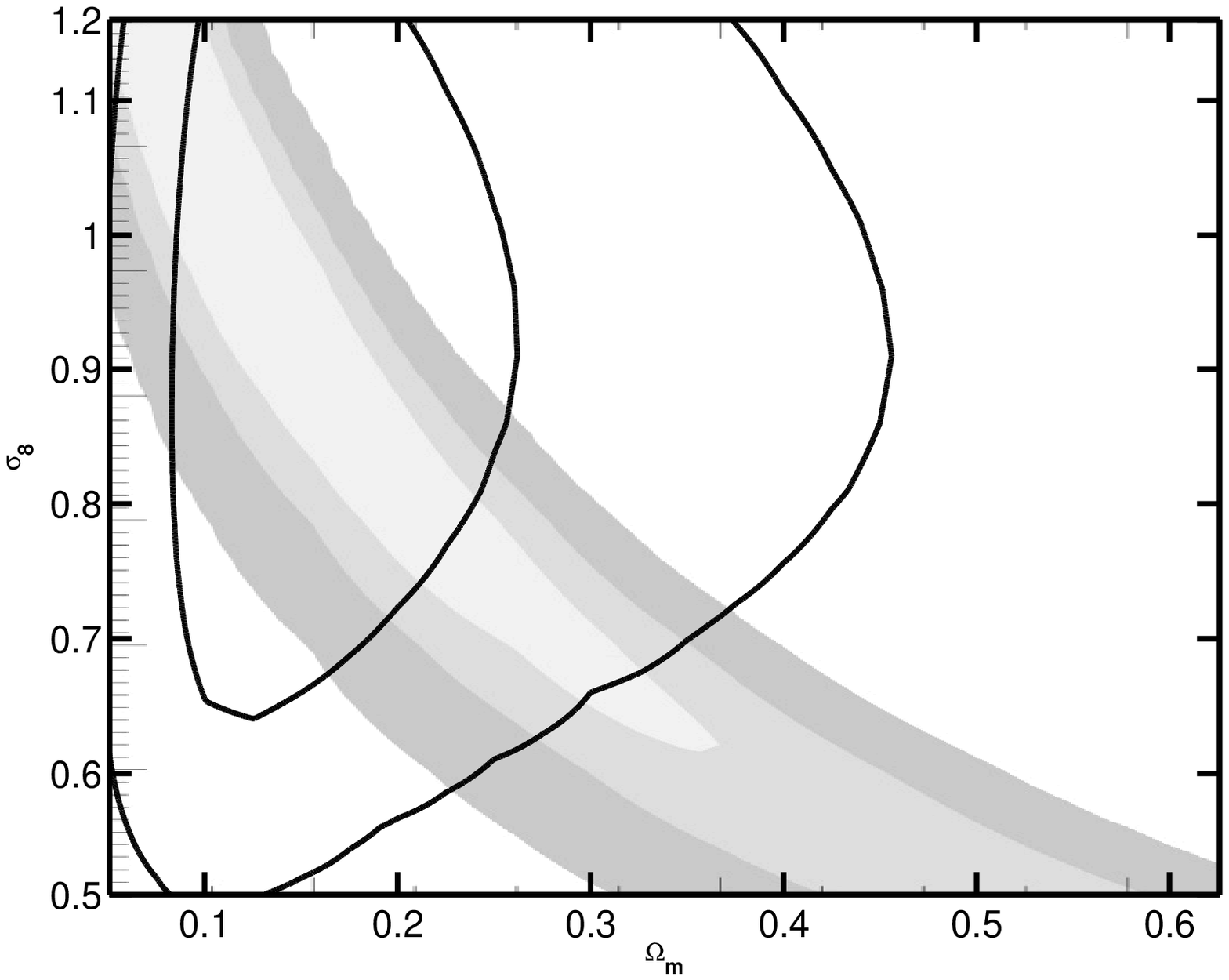,width=8cm,angle=0}
\epsfig{file=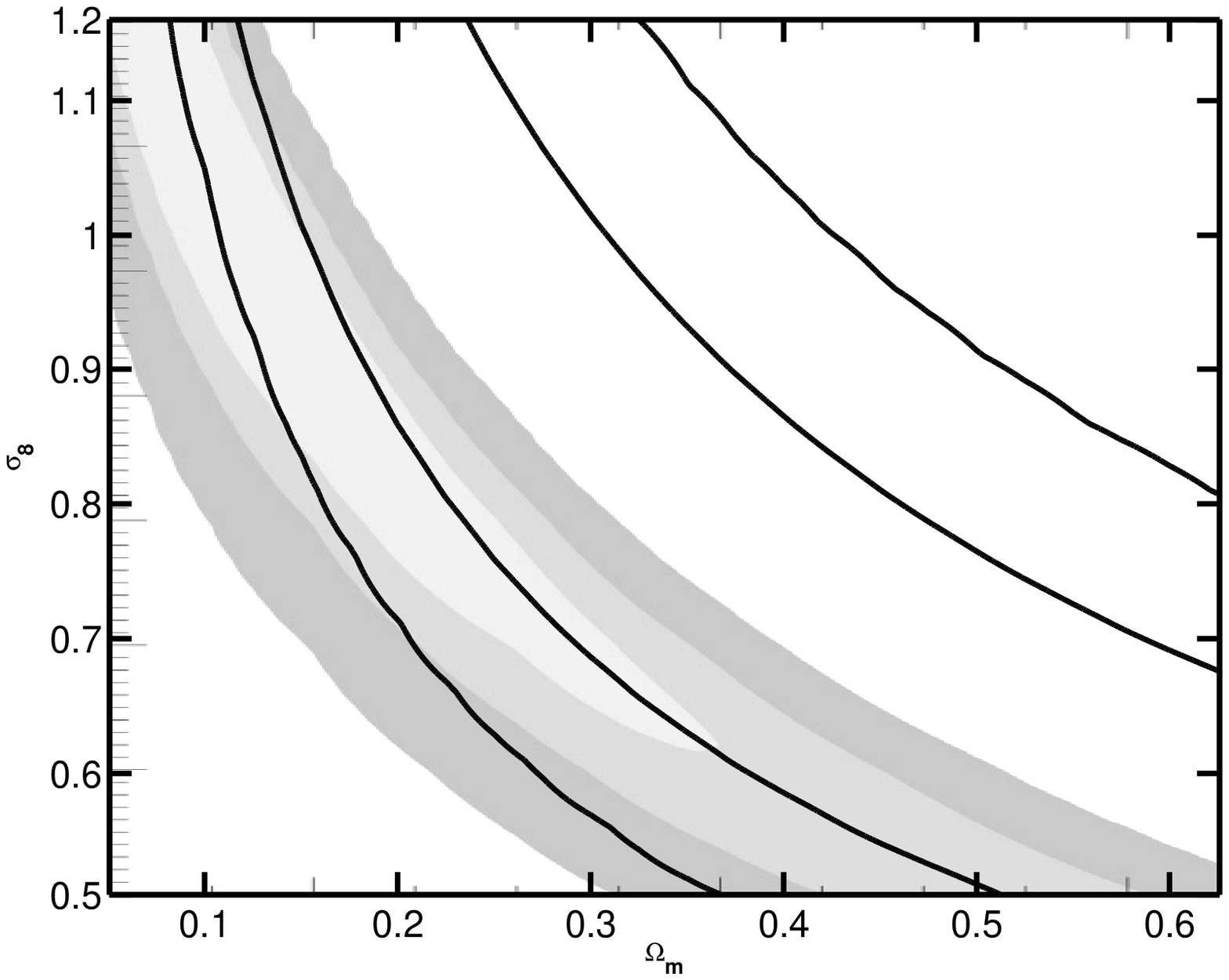,width=8cm,angle=0}
\caption{The thick contours in both panels are the 1 and 2$\sigma$ contours for the gridding method with a bin size of 25$h^{-1}$Mpc. 
The top panel shows the constraints when the global $\Omega_m$ is varied, and the bottom panel shows the constraints when just the $\Omega_m$ in the derivative of the growth function is varied.
The contours have been overlaid on the joint constraints on $\sigma_{8}$ and $\Omega_{m}$ from the 100 deg$^2$ weak lensing survey (grey scale contours) assuming a flat $\Lambda$CDM cosmology, depicting the 0.68, 0.95, and $0.99\%$ confidence levels.}
\label{varyom}
\end{figure}

With the optimum grid cell size of $L=25$Mpc/$h$ we constrain $\Omega_m$ and $\sigma_8$ simultaneously.   All cosmological parameters except for $\Omega_m$ and $\sigma_8$ were kept fixed as specified at the start of this section, and the binning has been accounted for by using Eq.~\ref{eq:corr}.  We vary $\Omega_m$  in two different senses.  First we vary the \textit{global} value of $\Omega_m$: the top panel  of Figure~\ref{varyom} shows the 1 and 2-$\sigma$ likelihood contours in the $\Omega_m$-$\sigma_8$ plane.  Then we set $\Omega_m=0.25$ in the power spectrum, and vary just the $\Omega_m$ in the derivative of the growth function $f(\Omega_m)$, which we shall now label $\Omega_m^{gro}$.  The 1 and 2-$\sigma$ likelihood contours of $\Omega_m^{gro}$-$\sigma_8$ are shown by the bottom panel of Figure~\ref{varyom}.  Both sets of contours have been overlaid on the joint constraints on $\sigma_\mathrm{8}$ and $\Omega_\mathrm{m}$ from the 100 deg$^2$ weak lensing survey \citep{ben} assuming a flat $\Lambda$CDM cosmology and adopting the nonlinear matter power spectrum of \cite{smith}.  The weak lensing contours depict the 0.68, 0.95, and $0.99\%$ confidence levels. The models are marginalised, over $h=0.72\pm0.08$, shear calibration bias, and the uncertainty in the redshift distribution.    

The resulting constraints on $\Omega_m$ and $\sigma_8$ are presented in rows 1, 2 and 3 of Table \ref{table:params}.  Column 1 are the constraints when marginalising over (the global) $\Omega_m$ in the case of $\sigma_8$ ($\sigma_8$ is marginalised over in the case of $\Omega_m$).  Column 2 is the same as column 1 except now these are the constraints where we have just varied $\Omega_m^{gro}$.  Column 3 is where we set the global $\Omega_m=0.25$ ($\sigma_8=0.9$ in the case of $\Omega_m$). Table \ref{table:params} shows that no improvement on the constraints is gained when the effect of $\Omega_m$ in the power spectrum is fixed.  All parameter constraints are consistent with $\Lambda$CDM and with WMAP5 to within 1-$\sigma$ confidence.

At the end of Section \ref{subsection:vcf} we mentioned the standard parameterisation of the linear growth function:
\begin{equation}
f\equiv \frac{d\ln \delta}{d\ln a}=\Omega_m^\gamma
\end{equation}
where the growth index $\gamma=6/11\simeq0.55$ for $\Lambda$CDM, and $\gamma\sim0.55$ for dark energy models with a slowly varying equation of state \citep{wangstein}.  In the context of modified gravity the growth index parameter can vary by as much as 30\%, with the prediction for DGP braneworld gravity \citep{DGP}: $\gamma = 0.69$ \citep{LC07}.  Therefore observational constraints on the growth index with better than 30\% uncertainty begin to have the power to discriminate between dark energy and modified gravity models.  

In this spirit we also present the likelihood contours in the $\Omega_m$-$\gamma$ plane in Figure \ref{varyomgamma}, to see how the SFI++ peculiar velocity data performs with this goal in mind.  Note that we only vary the global $\Omega_m$ simultaneously with $\gamma$. The contours shown are the 1 and 2-$\sigma$, the dark/blue solid line shows the $\Lambda$CDM growth index ($\gamma=0.55$) and the light/green dotted line shows the DGP growth index ($\gamma=0.69$).  After marginalising over $\Omega_m$, the 68 percent confidence constraint on the growth index is $\gamma=0.55^{+0.13}_{-0.14}$ which is consistent at 1-$\sigma$ with Einstein gravity, and just consistent at 1-$\sigma$ with DGP gravity; see also the 4th row in Table \ref{table:params}. There is no constraint in column 2 because we do not vary $\gamma$ separately with $\Omega_m^{gro}$.

\begin{table}
\caption{Parameter constraints from the SFI++ peculiar velocities.  Column 1 presents the 1-$\sigma$ constraint after marginalising over $\Omega_m$ ($\sigma_8$ in the case of $\Omega_m$).  
Column 2 is as column 1 except we have just varied $\Omega_m$ in $f(\Omega_m)$.
Column 3 presents the 1-$\sigma$ constraint after setting $\Omega_m=0.25$ ($\sigma_8=0.9$ in the case of $\Omega_m$).}
\begin{center}
\begin{tabular}[b]{|l|c|c|c|} 
\hline 
Parameter & Marg1 & Marg2 & $\Omega_m=0.25$ (or $\sigma_8=0.9$)\\
\hline 
$\sigma_8$             &$0.91^{+0.22}_{-0.18}$ & $0.56^{+0.48}_{-0.19}$ &$0.91^{+0.22}_{-0.16}$\\
$\Omega_m$          &$0.13^{+0.10}_{-0.04}$ & -                                          &$0.13^{+0.11}_{-0.03}$\\
$\Omega_m^{\tiny{gro}}$& -                                   & $0.10^{+0.49}_{-0.05}$ &$0.25^{+0.13}_{-0.08}$ \\
$\gamma$               &$0.55^{+0.13}_{-0.14}$ & -                                          &$0.53^{+0.15}_{-0.13}$\\
\hline
\end{tabular}
\end{center}
\label{table:params}
\end{table}

\section{Discussion}
\label{section:disc}

In this section we discuss our results in light of other recent constraints found in the literature.

\subsection{Comparison with other constraints}

We jointly constrained $\sigma_8$ and $\Omega_m$, shown in Figure \ref{varyom} where we have also plotted the constraints from the 100 deg$^2$ weak lensing survey. As can be seen from this figure, the weak lensing contours follow lines of constant $\sigma_8\Omega_m^{\gamma}$, the normalisation of the velocity signal.  The contours from the peculiar velocities in the top panel however (where we vary the global $\Omega_m$) are more independent of the value of $\Omega_m$. The joint effects of $\Omega_m$ in the normalisation and in the shape of the matter power spectrum partially cancel out to produce the flatter contours.

The bottom panel of Figure \ref{varyom} shows the $\sigma_8$ and $\Omega_m^{gro}$ 1 and 2-$\sigma$ contours, when just the $\Omega_m$ in $f(\Omega_m)$, i.e. the normalisation of the velocity signal divided by $\sigma_8$, is varied.  The shape of these contours now matches the shape of the weak lensing contours; it follows lines of constant $\sigma_8\Omega_m^{\gamma}$. This panel clearly illustrates how the different cosmological effects alter the shape of the contours.

After marginalising over (the global) $\Omega_m$ we found $\sigma_8=0.91\pm0.20$,  see Table \ref{table:params}.  This is consistent with the value of $\sigma_8$ estimated by \cite{GLS07} from SN1a data. \cite{GLS07} used the VCF to calculate the peculiar velocity field induced covariance of the luminosity distances and, given a set of cosmological parameters, fit this to the SN1a data.  They found $\sigma_8=0.79\pm0.22$ from a data set of 124 SN1a with a median redshift of $\bar{z}=0.024$, similar to the median redshift of SFI++.  It should  be noted that their constraint includes priors on $\Omega_bh^2$ and $h$, whereas we fix both $\Omega_b$ and $h$ to concordance values.

These results are seemingly in conflict with those found in \cite{wfh09}, who measured the bulk flow from several peculiar velocity surveys combined and found a bulk flow on large scales way in excess of what would be expected from $\Lambda$CDM.  Their combined sample, while it is a combination of many different peculiar velocity surveys, is dominated by SFI++. The bulk flow they measured from SFI++ data alone was also inconsistent with $\Lambda$CDM. In addition they computed confidence levels in the $\sigma_8$-$\Omega_mh^2$ plane for the observed bulk flow (shown in Figure 6 of their paper). They found that the WMAP 2-$\sigma$ confidence levels for these parameters were excluded at greater than 2-$\sigma$ confidence.  Fixing all parameters except $\sigma_8$ to central WMAP5 values they find $\sigma_8\sim1.7$, with 2-$\sigma$ and 3-$\sigma$ lower limits of 1.109 and 0.878 respectively, excluding the WMAP5 value of $\sigma_8$ at more than 3-$\sigma$.  This result is a remarkably different to our constraint on $\sigma_8$ after we fix all the other parameters to the concordance values stated at the start of Section \ref{section:res} (these are only marginally different from the central WMAP5 values).  We find $\sigma_8=0.91^{+0.22}_{-0.18}$ and although the peak of the likelihood is at a relatively high value of $\sigma_8$, the result is still consistent with WMAP5.

However our constraints are derived from a range of scales (the bulk flow is just the first moment of the flow on one scale) and these scales are mixed with one another, all contributing with different weights to the result.  It is likely that in our measurement more weight is given to the 25-50 $h^{-1}$Mpc scale range because the larger scales are drowned out due to the larger uncertainties at greater distances. The measurement made by \cite{wfh09} was based just on the bulk flow on a 50$h^{-1}$Mpc scale.  The inconsistencies between our results and theirs may just be due to the different sensitivities of the bulk flow statistic and the VCF.

As a check we perform a simple bootstrap test: we split the SFI++ data into two equally sized samples, one sample contains the galaxies with the nearest measured distances and the other the furthest.  We then repeat the analysis (constraining the global $\Omega_m$ and $\sigma_8$ simultaneously) on each sample separately.  We find no statistically significant difference between the two samples, and the marginalised constraints on $\sigma_8$ are as follows:  $\sigma_8=1.11^{+0.27}_{-0.23}$ (near sample), $\sigma_8=0.76^{+0.29}_{-0.27}$ (far sample).  In fact we find the central $\sigma_8$ value for the far sample, which probes mainly the larger scales, to be closer to the WMAP5 central value than the near sample: an opposite result to \cite{wfh09}, but given the size of the uncertainties, this is not significant.


\subsection{Comparison with projected 6dFGS performance}

In \cite{abate6df} forecasts were made for the performance of the Six Degree Field Galaxy Survey \citep[6dFGS,][]{6df} galaxy velocity survey.  \cite{abate6df} performed equivalent analysis to that done here on mock catalogues of the 6dFGS survey. They found the expected error on $\sigma_8$ after marginalising over $\Omega_m$ was approximately 16 percent, compared to our current estimate of 20 percent from SFI++ data.  This shows that the statistical improvement in this instance does not scale with the square root of the number of galaxies in the survey as might be expected.  This can be understood by looking at the redshift distributions of the 6dfGS mock catalogue and the SFI++ data: the median redshift of the 6dFGS mock is $\bar{z}=0.040$, a great deal larger than the SFI++ median redshift of $\bar{z}=0.019$.  With a larger median redshift 6dFGS will have larger statistical uncertainties on the peculiar velocities meaning less constraining power per galaxy than SFI++.

The actual statistical improvement on parameter constraints from 6dFGS over SFI++ will depend on the exact properties of the final data set.  The release of the 6dFGS peculiar velocities is expected shortly (Magoulas et al. and Springob et al, in prep). 6dFGS will perhaps be most important for investigating the results of \cite{kash} and \cite{wfh09}.  Because it samples such a large volume it will be possible to measure the bulk flow at scales between 50$h^{-1}$Mpc \cite[as measured by][]{wfh09} and 300$h^{-1}$Mpc \cite[as measured by][]{kash} thus
improving our understanding of the bulk flow and confirming if there is a discrepancy with $\Lambda$CDM.

\subsection{Modified gravity constraints}

\begin{figure}
\center
\epsfig{file=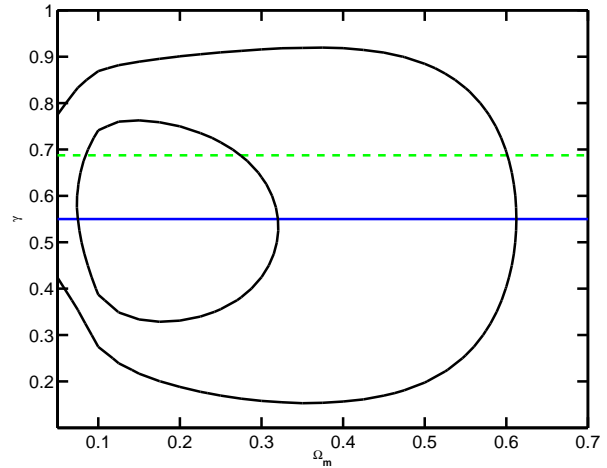,width=8cm,angle=0}
\caption{1 and 2$\sigma$ likelihood contours in the $\Omega_m$-$\gamma$ plane. The dark/blue horizontal solid line shows the $\Lambda$CDM growth index ($\gamma=0.55$) and the light/green horizontal dotted line shows the DGP growth index ($\gamma=0.69$).}
\label{varyomgamma}
\end{figure}

We also presented joint constraints on the growth index $\gamma$ and $\Omega_m$, and found $\gamma=0.55\pm0.14$ after marginalising over $\Omega_m$, see Figure \ref{varyomgamma} and Table \ref{table:params}. This value is consistent with Einstein gravity and just consistent with DGP gravity and 1-$\sigma$ confidence.
Other recent observational constraints of the growth index include \cite{rapetti08} and \cite{nesseris08}; we find our constraint to be consistent with these, and with a similar amount of statistical error. \cite{rapetti08} constrained $\gamma$ from the X-ray luminosity function and found $\gamma=0.51^{+0.16}_{-0.15}$ for the $\Lambda$CDM model.  Although they include extra data sets (CMB, SN1a, X-ray cluster gas-mass fractions) they also marginalise over a much larger parameter space, which includes allowances for systematic uncertainties, and so is a more rigorous analysis.  \cite{nesseris08} on the other hand constrain the growth function $f$ indirectly, mainly through observations of $\sigma_8(z)$ inferred from Ly-$\alpha$, and they find $\gamma=0.67^{+0.20}_{-0.17}$.

As the amount of data available for this type of analysis increases it will be interesting to see if the growth index reveals any possible departures from the $\Lambda$CDM/dark energy predicted value of $\gamma\simeq0.55$.

\section{conclusions}
\label{section:conc}

We presented cosmological parameter constraints using the SFI++ galaxy peculiar velocity data.  We performed the analysis by using the gridding method developed in \cite{abate6df}.  The gridding method is an efficient way to analyse the data given the large number of velocities in the survey, and without gaining biases on the cosmological parameters from the nonlinear growth of structure.  By comparing with a linear simulation we found the optimum scale on which to average the peculiar velocity data was $L=25$Mpc/$h$, and this scale is consistent with the optimum scale found in \cite{abate6df}.

To summarise the main points:
\begin{itemize}

\item Galaxy and SN1a peculiar velocities provide the only way to truly measure clustering at redshift zero.  They can provide a low redshift anchor for measuring the evolution of the growth of structure.

\item Peculiar velocities are sensitive to the derivative of the growth function, and can provide competitive constraints on parameters such as $\sigma_8$ and $\gamma$.  Therefore they are a useful tool for discriminating between $\Lambda$CDM and modified gravity models.

\item Our constraints from SFI++ peculiar velocities are consistent with many other constraints found in the literature and with $\Lambda$CDM.  However the constraints come from a mixed range of scales and necessarily will be dominated by the smaller scales which have the least uncertainties. Any departure from $\Lambda$CDM on larger scales ($>100h^{-1}$Mpc) will not be well tested by this data set.

\item The different degeneracy direction of constraints in the $\sigma_8$-$\Omega_m$ plane compared to e.g. weak lensing will also be useful in breaking degeneracies, see also \cite{GLS07}. 

\item Because peculiar velocities will be subject to different systematic effects to other probes such as cluster-counts, weak lensing or redshift distortions, they are an important complementary probe of cosmology.

\end{itemize}
In the near future much more peculiar velocity data will become available, boosting even more the importance of this kind of analysis. The new technique of measuring galaxy cluster peculiar velocities using the kSZ could soon be achieved in large samples by Sunyaev-Zeldo'vich (SZ) surveys such as the Atacama Cosmology Telescope \citep[ACT,][]{act} and the South Pole Telescope \citep[SPT,][]{spt}.  The kSZ is the temperature shift of CMB photons caused by a moving galaxy cluster.  It is proportional to the line-of-sight momentum of the cluster gas, so therefore the radial peculiar velocity of the cluster can be inferred.  The huge advantage of this method over using galaxies or SN1a is that the kSZ measurement is redshift independent.  This means we can probe the velocity field out to much larger redshifts, and with a greater accuracy than with galaxy peculiar velocity surveys, see \cite{bhattkosow} for forecasts on the possible constraint on $w$. Although measurement of the kSZ signal is still a difficult challenge observationally \citep{khc,diaferio}.

On a shorter timescale, an upcoming release of the Six Degree Field Galaxy Survey will contain around 10 000 galaxy peculiar velocities, an order of magnitude larger than most peculiar velocity surveys to date.  Furthermore there are ever growing SN1a samples \citep[e.g.][]{jrk}, and current and upcoming surveys such as Skymapper \citep{skymap} and GAIA \citep{gaia} should push the sample size of SN1a well into the thousands.

\section{Acknowledgements} We thank Sarah Bridle, Hume Feldman, Andrew Jaffe, Ofer Lahav, Karen Masters and Christopher Springob for their many helpful comments and discussion.

\end{document}